\documentclass[letterpaper,11pt]{article}
\usepackage[DIV12]{typearea}
\usepackage{longtable}
\usepackage{booktabs}
\usepackage{mathtools}
\usepackage{amsmath}
\usepackage{amssymb}
\usepackage{bbm}
\usepackage{dsfont}
\usepackage[utf8]{inputenc}
\usepackage{pdflscape}
\usepackage{xspace}
\usepackage[caption=false]{subfig}
\usepackage{nicefrac}
\usepackage{graphicx}
\usepackage{cite}
\usepackage{nccfoots}
\usepackage{cancel}
\usepackage[small]{caption2}
\usepackage{multirow}
\usepackage[normalem]{ulem}

\newcommand{\vev}[1]{\ensuremath{\langle{#1}\rangle}}

\newcommand{\rep}[1]{\ensuremath\boldsymbol{#1}}
\newcommand{\crep}[1]{\ensuremath\bar{\boldsymbol{#1}}}

\newcommand{\Z}[1]{\ensuremath{\mathbbm{Z}_{#1}}} % z_N ->\Z{N}

\newcommand{\SL}[1]{\ensuremath{\mathrm{SL}(#1)}}
\newcommand{\PSL}[1]{\ensuremath{\mathrm{PSL}(#1)}}

\newcommand{\I}{\mathrm{i}}
\newcommand{\e}{\mathrm{e}}
\newcommand{\Id}{\mathds{1}}

\newcommand{\nphantom}[1]{\sbox0{#1}\hspace{-\the\wd0}}

\usepackage[nolist]{acronym}
\usepackage[textsize=tiny,backgroundcolor=yellow]{todonotes}  %eliminate later

\advance \headheight by 3.0truept       % for 12pt mandatory...
\usepackage{xcolor}
\definecolor{darkgreen}{HTML}{109930}
\usepackage[pdftex,hidelinks,colorlinks=true,citecolor=darkgreen]{hyperref}%,linkcolor=blue,citecolor=blue
\hypersetup{
    pdftitle = {Demystifying stringy miracles with eclectic flavor symmetries},
    pdfauthor = {Knapp-Perez, Liu, Nilles, Ramos-Sanchez}
}

\usepackage[noabbrev]{cleveref}% wants to be loaded last, use \Cref
\def\mytitle{Demystifying stringy miracles with eclectic flavor symmetries}
\title{\mytitle}
\numberwithin{equation}{section}
\numberwithin{figure}{section}
\numberwithin{table}{section}

\allowdisplaybreaks
\usepackage{braket}%Dirac Notation

\begin{document}

%%%%%%%%%%%%%%%%%%%%%%%%%%%%%%%%%%%%%%%%%%%%%%%%%%%%%%%%%%%%%%%%%%%%%%%%%%
%  Title
%%%%%%%%%%%%%%%%%%%%%%%%%%%%%%%%%%%%%%%%%%%%%%%%%%%%%%%%%%%%%%%%%%%%%%%%%%
\begin{titlepage}
\vspace*{2cm}

\begin{center}
{\Large\bfseries\mytitle}

\vspace{1cm}

\renewcommand*{\thefootnote}{\fnsymbol{footnote}}

\textbf{%
V. Knapp--P\'erez$^{a,}$\footnote{vknapppe@uci.edu},
Xiang--Gan Liu$^{a,}$\footnote{xianggal@uci.edu},
Hans Peter Nilles$^{b,}$\footnote{nilles@uni-bonn.de}\\
and
Sa\'ul Ramos-S\'anchez$^{c,}$\footnote{ramos@fisica.unam.mx}
}
\\[8mm]
\textit{$^a$\small~Department of Physics and Astronomy, University of California, Irvine, CA 92697-4575 USA}
\\[5mm]
\textit{$^b$\small Bethe Center for Theoretical Physics and Physikalisches Institut der Universit\"at Bonn,\\ Nussallee 12, 53115 Bonn, Germany}
\\[5mm]
\textit{$^c$\small Instituto de F\'isica, Universidad Nacional Aut\'onoma de M\'exico, Cd.\ de M\'exico C.P.\ 04510, M\'exico}
\end{center}

\vspace*{1cm}

\begin{abstract}

Effective field theories arising from string compactifications are subject to
constraints originating from the duality transformations of string theory.
Interpreting these so-called selection rules in terms of conventional symmetries
has remained challenging. We show that particular selection rules in heterotic
orbifolds can be explained from a subtle interplay between modular and traditional
flavor symmetries within the eclectic flavor framework.
\end{abstract}

\vspace*{1cm}
\end{titlepage}
\renewcommand*{\thefootnote}{\arabic{footnote}}
\setcounter{footnote}{0}

%%%%%%%%%%%%%%%%%%%%%%%%%%%%%%%%%%%%%%%%%%%%%%%%%%%%%%%%%%%%%%%%%%%%%%%%%%
%  Intro
%%%%%%%%%%%%%%%%%%%%%%%%%%%%%%%%%%%%%%%%%%%%%%%%%%%%%%%%%%%%%%%%%%%%%%%%%%
\section{Introduction}

Symmetries play an important role in particle physics. The
absence of certain couplings in an effective field theory
is usually interpreted as a consequence of an underlying symmetry.
As a ``folklore theorem,'' one can state that any coupling that is not
explicitly forbidden by a symmetry is expected to appear
in the action. This idea is frequently formulated as a so-called
``naturalness criterion,'' which aims to explain the absence
(or smallness) of certain couplings.

Nonetheless, we are sometimes confronted with a situation
where certain selection rules in a theory apparently cannot
be explained by a symmetry and thus appear ``miraculous''.
Very often, however, a more careful inspection of the
situations reveals the existence of a ``hidden symmetry''
that explains the selection rule, restores the naturalness
argument and demystifies the ``miracle''.

In the present paper, we discuss a stringy selection rule
which for more than 30 years has resisted an explanation in
terms of a symmetry argument in the 4-dimensional low-energy
effective action. It originates in \ac{CFT}
correlators in orbifold compactifications of heterotic
string theory~\cite{Hamidi:1986vh,Dixon:1986qv}, most notably
in the \Z3 orbifold. There, one encounters the so-called Rule 4~\cite{Font:1988tp},
a selection rule which states that certain
couplings of twisted fields vanish when all the fields
are located at the same fixed point. It was argued~\cite{Font:1988nc} 
that such a selection rule cannot be explained as a consequence of a symmetry
of the effective action. As a main result of the present
paper, we show that Rule 4 can indeed be explained
as a consequence of a ``hidden'' symmetry. The argument
is based on so-called modular and eclectic symmetries~\cite{Nilles:2020nnc}
that originate from duality transformations in string
theory. A priori, these are not conventional symmetries
as they map a theory not necessarily to itself, but rather to
its dual. In many cases, however, they lead to selection
rules in the low-energy effective theory that can be
understood in the eclectic symmetry framework~\cite{Nilles:2020nnc}.
This scheme combines discrete modular symmetries (that act
nontrivially on a modulus field) with traditional flavor
symmetries (which act as conventional discrete
symmetries that leave the modulus invariant). They
necessarily appear together and should not be considered
in isolation, as these discrete modular symmetries are connected
to the group of the outer automorphisms of the traditional flavor
symmetry. At the heart of this construction is a hybrid
$\Z2^\mathrm{hybrid}$ symmetry in the modular group $\SL{2,\Z{}}$
(not contained in $\PSL{2,\Z{}}$), which is intrinsically modular
but does not transform the modulus and can, thus, be regarded as a traditional flavor symmetry too. This describes the hybrid nature of $\Z2^\mathrm{hybrid}$.

As Rule 4 is formulated in the framework of the \Z3 orbifold,
we shall exclusively consider this case here. Our results
concerning the role of the hybrid $\Z2^\mathrm{hybrid}$ symmetry and the
eclectic scheme, however, shall also be relevant for more general
cases to be discussed in future work. In the $\mathds T^2/\Z3$ orbifold
we have the traditional flavor symmetry $\Delta(54)$ and the
modular flavor symmetry $T'$. The hybrid $\Z2^\mathrm{hybrid}$ is contained in
both of them. It extends $\Delta(27)$ to $\Delta(54)$ and $A_4$
to its double cover\footnote{$T'$ and $A_4$ are finite modular groups
arising from $\SL{2,\Z{}}$ and $\PSL{2,\Z{}}=\SL{2,\Z{}}/\Z2$, respectively.} $T'$.
In the following we shall show that Rule 4 can be explained in a rather
subtle way via the interplay of modular and traditional flavor
symmetry.

We build our discussion on the earlier observation~\cite{Nilles:2020kgo}
that Rule 4 was consistent with the appearance of specific
trilinear couplings of twisted fields. Here we reanalyze this
result in detail and extract the basic reason for this fact.
This insight will then allow us to prove Rule 4 in the general
case in terms of the eclectic flavor scheme. At the heart of
this proof is the hybrid $\Z2^\mathrm{hybrid}$ symmetry in combination with
a $\Z3^\mathrm{rot}$ symmetry that rotates the three fixed points of the \Z3 orbifold.
It is a subgroup of $\Delta(27)$ in $\Delta(54)$.
$\Z2^\mathrm{hybrid}$ and $\Z3^\mathrm{rot}$ combine to an $S_3^R$ non-Abelian
$R$-symmetry.\footnote{A geometric construction of the $S_3$ group is given
in detail in Ref.~\cite{Baur:2019kwi}.} This shows that the eclectic scheme
explains Rule 4 and demystifies what was thought to be a
``stringy miracle''. We thus see that, apart from understanding
Rule 4, we reveal a subtle interplay between modular and
traditional flavor symmetries that might have important
consequences for selection rules beyond the case of the
\Z3 orbifold.

The paper is organized as follows. \Cref{sec:rule4} explains Rule 4 as
it was discussed in the framework of orbifold conformal field
theory. In \cref{sec:Z3eclectic} we present the concept of the eclectic
flavor symmetry for the $\mathds T^2/\Z3$ orbifold, the appearance of
the hybrid $\Z2^\mathrm{hybrid}$ as well as the representations of the twisted
fields under $\Delta(54)$ and $T'$. \Cref{sec:trilinear} recalls the
explicit calculation of the trilinear couplings as provided
earlier~\cite{Nilles:2020kgo} and analyzes the implications of $T'$ and
$\Delta(54)$ separately. We will show that there are two dual
explanations of Rule 4 starting either from $\Delta(54)$ or
$T'$. Armed with these observations, we give the general proof
of Rule 4 in these dual ways starting from either the traditional or the
modular symmetry. We shall see that in both cases the
appearance of $\Z2^\mathrm{hybrid}$ and $\Z3^\mathrm{rot}$ are crucial. We
also point out a loophole in the earlier discussion~\cite{Font:1988nc}  where
one did not consider the possible role of non-Abelian discrete
$R$-symmetries. \Cref{sec:higherOrder} discusses some immediate
consequences of the symmetry explanation of Rule 4 in the
presence of Wilson lines. Wilson lines typically break the
degeneracy of the fixed points to allow for realistic
spectra, as discussed in detail in Ref.~\cite{Ibanez:1986tp}. Since the
$\Z3^\mathrm{rot}$ is responsible for this degeneracy, it is broken
in the presence of Wilson lines. As this symmetry is crucial
for the proof of Rule 4, one would thus expect that such a
rule is no longer valid once a Wilson line is switched on. Some consequences
of these observations are briefly discussed in \cref{sec:consequences}.
\Cref{sec:conclusions} provides a summary and outlook on possible future
investigations.

%%%%%%%%%%%%%%%%%%%%%%%%%%%%%%%%%%%%%%%%%%%%%%%%%%%%%%%%%%%%%%%%%%%%%%%%%%
%  Rule 4
%%%%%%%%%%%%%%%%%%%%%%%%%%%%%%%%%%%%%%%%%%%%%%%%%%%%%%%%%%%%%%%%%%%%%%%%%%

\section{Selection Rule 4 in heterotic orbifolds}
\label{sec:rule4}

Couplings arising from orbifold compactification can be computed within the context of \ac{CFT}~\cite{Hamidi:1986vh,Dixon:1986qv}. They lead to selection rules that have been formulated in Ref.~\cite{Font:1988tp}. Of particular interest is a selection rule known as Rule 4, as discussed in Ref.~\cite{Font:1988nc} and refined in Ref.~\cite{Kobayashi:2011cw}. Rule 4 states that certain local correlation functions of twisted fields vanish when all fields are located at the same fixed point. This has been explicitly discussed in the framework of the \Z3 orbifold in Refs.~\cite{Font:1988tp,Font:1988nc}.

Relevant for this rule is the presence of oscillator modes and derivatives in the corresponding vertex operators of the CFT~\cite{Font:1988nc}. When computing $n$-point Yukawa couplings with \ac{CFT} methods, the picture-changing mechanism for correlation functions leads to $n-3$ derivatives. In the $\mathds T^2/\Z3$ case, Rule 4 states that
couplings of twisted fields {\it at the same fixed point} are only allowed if the sum $\Sigma$ of the number of these derivatives and the number of oscillator modes is equal to $0 \bmod 6$.

This rule appears somewhat peculiar as there is no such rule for similar couplings of the same twisted fields located at different fixed points. This leads to the question whether such a selection rule can be understood in terms of the symmetries of the low energy effective 4-dimensional quantum field theory. A first analysis of the question~\cite{Font:1988nc} led to the conclusion that conventional symmetries were not able to explain Rule 4. This would mean that certain selection rules of string theory might not be understandable through symmetries of the low energy effective field theoretical approximation. In the following we shall try to clarify this question.

%%%%%%%%%%%%%%%%%%%%%%%%%%%%%%%%%%%%%%%%%%%%%%%%%%%%%%%%%%%%%%%%%%%%%%%%%%
%  Eclectic scheme
%%%%%%%%%%%%%%%%%%%%%%%%%%%%%%%%%%%%%%%%%%%%%%%%%%%%%%%%%%%%%%%%%%%%%%%%%%

\section{Eclectic scheme in $\mathds T^2/\Z3$}
\label{sec:Z3eclectic}

\subsection{The origin of the eclectic symmetries}

In the $\mathds T^2/\Z3$ orbifold of the heterotic string, there are various symmetries
of the associated low-energy effective field theory that are explained as symmetries of the
toroidal orbifold compactification of the extra dimensions of a heterotic string in the Narain formalism~\cite{GrootNibbelink:2017usl}.
First, a $\mathds T^2$ of the heterotic string is characterized by two moduli: a complex structure $U$
and a K\"ahler modulus $T$. The complex structure is geometrically stabilized at $\vev{U} = \omega := \e^{\nicefrac{2\pi\I}{3}}$,
so that the two-torus  $\mathds T^2$ exhibits a \Z3 rotational symmetry compatible with the \Z3 orbifold.
In this case, the modular group $\SL{2,\Z{}}_U$ associated with the complex structure $U$ is broken.
Further, one can identify two unbroken rotational outer automorphisms of the $\mathds T^2/\Z3$ Narain
space group corresponding to the generators $\mathrm S$ and $\mathrm T$ of the modular group $\SL{2,\Z{}}_T$
for the K\"ahler modulus. These generators are subject to the constraints
\begin{equation}
\label{eq:SL2Zcond}
  \mathrm S^4 ~=~ \Id ~=~ (\mathrm{ST})^3\qquad \text{and} \qquad \mathrm S^2 \mathrm T ~=~ \mathrm T \,\mathrm S^2\,.
\end{equation}
Further, they act nontrivially on matter fields~\cite{Lauer:1989ax,Lerche:1989cs,Chun:1989se}, yielding modular flavor symmetries.

On the other hand, there are two translational outer automorphisms of the $\mathds T^2/\Z3$ Narain space group, denoted
$\mathrm A$ and $\mathrm B$. Since translations leave all moduli invariant and do transform matter fields of the orbifold,
they correspond to traditional flavor symmetries. Additionally, there exists a rotational outer automorphism
$\mathrm{C}$ of the Narain space group, which leaves the moduli invariant and acts nontrivially on matter fields.
Hence, $\mathrm C$ qualifies also as a traditional flavor symmetry. It turns out that, at the level of outer
automorphisms of the Narain space group, one finds that
\begin{equation}
\label{eq:hybridId}
  \mathrm C ~=~ \mathrm S^2\,,
\end{equation}
showing that the generator $\mathrm C$ builds both a traditional and a modular flavor symmetry, i.e.\ it is
a {\it hybrid symmetry generator}. Due to \Cref{eq:SL2Zcond}, $\mathrm S^2$ describes a $\Z2^\mathrm{hybrid}$ symmetry.

\subsection{Matter and the eclectic flavor group}
\begin{figure}[t!]
    \centering
    \includegraphics{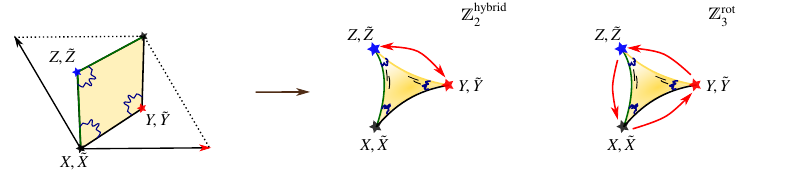}
    \caption{\label{fig:Z3orbifold} The $\mathds{T}^2/\Z3$ orbifold and two of its symmetries
    in the absence of Wilson lines.}
\end{figure}

The action of the various generators on matter fields depends on their localization in the extra dimensions:
bulk strings build singlets while twisted strings, localized at the three fixed points of the
$\mathds T^2/\Z3$ orbifold (see \Cref{fig:Z3orbifold}), transform as (reducible) triplets. Under modular transformations
matter fields carry a (rational) modular weight $n$ and transform as
\begin{equation}
\label{eq:ModularTransformationOfFields}
\Phi_n ~\xmapsto{~\gamma~}~ (c\,T+d)^n\,\rho_{\rep{s}}(\gamma)\,\Phi_n\,,
\qquad \gamma~=\begin{pmatrix}a&b\\c&d\end{pmatrix}~\in~\SL{2,\Z{}}_T\,,
\end{equation}
where $(cT+d)^n$ is called automorphy factor, and the $\SL{2,\Z{}}_T$ generators can be written as
\begin{equation}
\mathrm S ~=\begin{pmatrix} 0&1\\-1&0\end{pmatrix}\qquad\text{and}\qquad
\mathrm T ~=\begin{pmatrix} 1&1\\0&1\end{pmatrix}.
\end{equation}
Further, bulk states transform with the representations $\rho_{\rep1}(\mathrm S)=\rho_{\rep1}(\mathrm T)=1$
and have the modular weights $n=0$ or $n=-1$. The three twisted states located at three fixed points
of the orbifold are collected in two kinds of multiplets:
\begin{subequations}\label{eq:ThetaTwistedMatter}
\begin{eqnarray}
\Phi_{-\nicefrac{2}{3}} & = & (X, Y, Z)^\mathrm{T}                         \quad\mathrm{without\ oscillator\ excitations}\;,\\
\Phi_{-\nicefrac{5}{3}} & = & (\tilde{X}, \tilde{Y}, \tilde{Z})^\mathrm{T} \quad\mathrm{with\ one\ holomorphic\ oscillator\ excitation}\;,
\end{eqnarray}
\end{subequations}
which transform under $\mathrm S$ and $\mathrm T$ according to
\begin{equation}
\label{eq:SandT}
  \rho(\mathrm S) = \frac{\I}{\sqrt3}\begin{pmatrix}1&1&1\\1&\omega^2&\omega\\1&\omega&\omega^2\end{pmatrix},
  \qquad
  \rho(\mathrm T) = \begin{pmatrix}\omega^2&0&0\\0&1&0\\0&0&1\end{pmatrix}.
\end{equation}
This implies that $\mathrm S$ and $\mathrm T$ act on matter fields as a \Z4 and a \Z3 symmetry, respectively.
Interestingly, these modular transformations build the $\rep2'\oplus\rep1$ irreducible representations of the finite modular
group\footnote{$[24,3]$ corresponds to the GAP notation~\cite{GAP4},
where the first number is the order and the latter only a counter.} $\Gamma_3'\cong T' \cong \SL{2,\Z3}\cong[24,3]$.
According to the irreducible basis in \Cref{tab:irrepTprime}, the fields in \Cref{eq:ThetaTwistedMatter} can be arranged as
\begin{equation}
\label{eq:2prime+1}
  \rep2': \begin{pmatrix}\tfrac{1}{\sqrt2}(Y+Z)\\-X \end{pmatrix},~\begin{pmatrix}\tfrac{1}{\sqrt2}(\tilde Y+\tilde Z)\\-\tilde X \end{pmatrix}
  \qquad\text{and}\qquad
  \rep1: \frac{1}{\sqrt2}(Y-Z),~\frac{1}{\sqrt2}(\tilde Y-\tilde Z).
\end{equation}

Under the generators of the traditional flavor symmetry, matter states $\Phi_{-\nicefrac23}$
build the $\Delta(54)$ triplet representation $\rep3_2$ while the multiplet $\Phi_{-\nicefrac53}$ builds
the representation $\rep3_1$ of $\Delta(54)$. The representation matrices of $\rep3_2$ and $\rep3_1$ are given in terms of
\begin{equation}
\label{eq:ABandC}
 \rho(\mathrm A) = \begin{pmatrix} 0&1&0\\0&0&1\\1&0&0\end{pmatrix},\quad
 \rho(\mathrm B) = \begin{pmatrix} 1&0&0\\0&\omega&0\\0&0&\omega^2\end{pmatrix},\quad
 \rho(\mathrm C) = \begin{pmatrix} 1&0&0\\0&0&1\\0&1&0\end{pmatrix}\;.
\end{equation}
In particular, $\rho_{\rep3_1}(\mathrm A)=\rho_{\rep3_2}(\mathrm A)=\rho(\mathrm A)$,
$\rho_{\rep3_1}(\mathrm B)=\rho_{\rep3_2}(\mathrm B)=\rho(\mathrm B)$
and $\rho_{\rep3_1}(\mathrm C)=-\rho_{\rep3_2}(\mathrm C)=\rho(\mathrm C)$. We observe that
$\rho(\mathrm A)$ represents the order-3 cyclic symmetry $\Z3^\mathrm{rot}$ that exchanges
all three matter states of the twisted multiplets $\Phi_n$, as illustrated in \Cref{fig:Z3orbifold}. Similarly,
the representation $\rho(\mathrm C)$ of the hybrid symmetry generator builds the $\Z2^\mathrm{hybrid}$,
which swaps $Y\leftrightarrow Z$ and $\tilde Y\leftrightarrow\tilde Z$.

The generators $\mathrm A$ and $\mathrm B$ yield the non-Abelian discrete group $\Delta(27)$. The inclusion of
$\mathrm{C}$ enhances the traditional flavor group to $\Delta(54)$, which is recognized as the traditional flavor symmetry of the
$\mathds T^2/\Z3$ orbifold~\cite{Kobayashi:2006wq}. Notably, $\mathrm{C}$ amounts to a $\pi$ rotation in the compact
dimensions~\cite{Baur:2019kwi}, making it a remnant of the higher-dimensional Lorentz symmetry. As a result, $\Delta(54)$ can be
interpreted as a non-Abelian discrete $R$-symmetry of $\mathcal N=1$ supersymmetry~\cite{Chen:2013dpa},
under which the superpotential transforms as a nontrivial singlet $\rep1'$ of $\Delta(54)$~\cite{Baur:2019iai}.
Further, we note that the traditional generators $\mathrm A$ and $\mathrm C$ build a non-Abelian
discrete $S_3^R\cong\Z3^\mathrm{rot}\rtimes\Z2^\mathrm{hybrid}$ symmetry.

Together, the traditional and modular symmetries build the eclectic group
$\Omega(1)\cong\Delta(54)\cup T'\cong [648,533]$, whose generators are given in \Cref{tab:Z3FlavorGroups}.
Under the different components of the group, all matter fields of the $\mathds T^2/\Z3$ orbifold transform
as described before and summarized in \Cref{tab:Representations}.

\begin{table}[t!]
\centering
\begin{tabular}{|c|c||c|c|c|c|c|}
\hline
\multicolumn{2}{|c||}{nature}                & outer automorphism       & \multicolumn{4}{c|}{\multirow{2}{*}{flavor groups}} \\
\multicolumn{2}{|c||}{of symmetry}           & of Narain space group    & \multicolumn{4}{c|}{}\\
\hline
\hline
\parbox[t]{3mm}{\multirow{5}{*}{\rotatebox[origin=c]{90}{eclectic}}} &\multirow{2}{*}{modular}  & rotation $\mathrm{S}~\in~\SL{2,\Z{}}_T$ & $\Z{4}$      & \multicolumn{2}{c|}{\multirow{2}{*}{$T'$}} &\multirow{5}{*}{$\Omega(1)$}\\
        &                          & rotation $\mathrm{T}~\in~\SL{2,\Z{}}_T$ & $\Z{3}$      & \multicolumn{2}{c|}{}                      & \\
\cline{2-6}
        & \multirow{2}{*}{traditional flavor} & translation $\mathrm{A}$                & $\Z{3}^\mathrm{rot}$      & \multirow{2}{*}{$\Delta(27)$} & \multirow{3}{*}{$\Delta(54)$} & \\
        &                                     & translation $\mathrm{B}$                & $\Z{3}^{\mathrm{B}}$        &                               & & \\
\cline{2-5}
        & hybrid \Z2                          & rotation $\mathrm{C}=\mathrm{S}^2~\in~\SL{2,\Z{}}_T$ & \multicolumn{2}{c|}{$\Z2^{\mathrm{hybrid}^{\phantom{A}}}$}  & & \\
\hline
\end{tabular}

\caption{\label{tab:Z3FlavorGroups}
Various symmetries of the eclectic flavor group $\Omega(1)\cong\Delta(54)\cup T'\cong [648, 533]$ for a $\mathds{T}^2/\Z{3}$ orbifold and their origin.
The order-3 translational outer automorphisms $\mathrm{A}$ and $\mathrm{B}$ generate $\Delta(27)$.
The rotational outer automorphism $\mathrm{C}$ is special as it belongs to both, the traditional flavor symmetry $\Delta(54)$ and the finite modular symmetry $T'$.
This table has been adapted from~\cite{Nilles:2020tdp}.}
\end{table}

\begin{table}[ht!]
\centering
\begin{tabular}{|c||c||c|c|c|c||c|c|c|c|}
\hline
\multirow{3}{*}{sector} &\!\!matter\!\!& \multicolumn{8}{c|}{eclectic flavor group $\Omega(1)$}\\ \cline{3-10}
                        &fields                                    & \multicolumn{4}{c||}{modular $T'$ subgroup} & \multicolumn{4}{c|}{traditional $\Delta(54)$ subgroup}\\
                        &$\Phi_n$                                  & \!\!irrep $\rep{s}$\!\! & $\rho_{\rep{s}}(\mathrm{S})$ & $\rho_{\rep{s}}(\mathrm{T})$ & $n$ & \!\!irrep $\rep{r}$\!\! & $\rho_{\rep{r}}(\mathrm{A})$ & $\rho_{\rep{r}}(\mathrm{B})$ & $\rho_{\rep{r}}(\mathrm{C})$ \\
\hline
\hline
bulk      & $\Phi_{\text{\tiny 0}}$   & $\rep1$             & $1$                   & $1$                   & $0$               & $\rep1$   & $1$               & $1$                   & $+1$               \\
          & $\Phi_{\text{\tiny $-1$}}$& $\rep1$             & $1$                   & $1$                   & $-1$              & $\rep1'$  & $1$               & $1$                   & $-1$               \\
\hline
$\theta$  & $\Phi_{\nicefrac{-2}{3}}$ & $\rep2'\oplus\rep1$ & $\rho(\mathrm{S})$    & $\rho(\mathrm{T})$    & $\nicefrac{-2}{3}$& $\rep3_2$ & $\rho(\mathrm{A})$& $\rho(\mathrm{B})$    & $-\rho(\mathrm{C})$\\
          & $\Phi_{\nicefrac{-5}{3}}$& $\rep2'\oplus\rep1$ & $\rho(\mathrm{S})$    & $\rho(\mathrm{T})$    & $\nicefrac{-5}{3}$& $\rep3_1$ & $\rho(\mathrm{A})$& $\rho(\mathrm{B})$    & $+\rho(\mathrm{C})$\\
\hline
super-    & \multirow{2}{*}{$\mathcal{W}$} & \multirow{2}{*}{$\rep1$} & \multirow{2}{*}{$1$} & \multirow{2}{*}{$1$} & \multirow{2}{*}{$-1$} & \multirow{2}{*}{$\rep1'$} & \multirow{2}{*}{$1$} & \multirow{2}{*}{$1$} & \multirow{2}{*}{$-1$} \\
\!\!potential\!\! & & & & & & & & & \\
\hline
\end{tabular}
\caption{\label{tab:Representations}
$T'$ and $\Delta(54)$ representations of (massless) matter fields $\Phi_n$
with modular weights $n$ in the untwisted and first twisted sectors of $\mathds{T}^2/\Z3$
orbifolds~\cite{Baur:2019kwi,Baur:2019iai}. $T'$ and $\Delta(54)$ combine
nontrivially to the $\Omega(1) \cong [648, 533]$ eclectic flavor group~\cite{Nilles:2020nnc},
generated by $\rho_{\rep{s}}(\mathrm{S})$, $\rho_{\rep{s}}(\mathrm{T})$,
$\rho_{\rep{r}}(\mathrm{A})$ and $\rho_{\rep{r}}(\mathrm{B})$. For $\rho_{\rep{r}}(\mathrm{C})$,
both $\mathrm{C}=\mathrm{S}^2$ and the modular weight $n$ are important, as discussed in \Cref{eq:C=S2rep}.
Table adapted from~\cite{Nilles:2020kgo}.}
\end{table}

Recall from \Cref{eq:hybridId} that $\mathrm C=\mathrm S^2$ and that this generator belongs
to both the traditional and modular parts of $\Omega(1)$, the eclectic flavor group.\footnote{This $\Z2^{\mathrm{hybrid}}$ overlap in $\Omega(1)\cong\Delta(54)\cup T'$ is precisely what the symbol $\cup$ implies. In principle, a more accurate group structure is $\Omega(1)\cong\Delta(27)\rtimes T'$, or equivalently $\Omega(1)\cong\Delta(54)\mathbin{\boldsymbol{.}}A_4$ , cf.~\cite[Appendix B.2]{Li:2025bsr}.} Hence, from \Cref{eq:ModularTransformationOfFields}, we obtain that
\begin{equation}
\label{eq:C=S2rep}
    \rho_{\rep r}(\mathrm C) ~:=~ (-1)^{n} \left(\rho_{\rep s}(\mathrm S)\right)^2\;.
\end{equation}
For example, the bulk fields $\Phi_0$ and $\Phi_{-1}$, which are trivial singlets under the modular group $T'$,
exhibit a sign difference in their $\Delta(54)$ matrix representation $\rho_{\rep r}(\mathrm C)$. In contrast,
for the twisted sector, the presence of fractional modular weights makes $(-1)^n$ multivalued, taking values
in $\{1, \omega, \omega^2\}$. This factor gives rise to a traditional flavor symmetry associated with the point
group $\Z{3}^{\mathrm{(PG)}}$. Consequently, within the eclectic framework, the modular weights and the
traditional flavor symmetry representation must be selected in a consistent way.

\subsection[Yukawa couplings in the T2/Z3 orbifold]{Yukawa couplings in the $\mathds T^2/\Z3$ orbifold}
\label{subsec:TprimerForms}

Since $\Gamma_3'\cong T'$ is the modular flavor symmetry of the eclectic scheme of the $\mathds T^2/\Z3$ orbifold,
all Yukawa couplings are given by $T'$ \acp{VVMF}.
All \acp{VVMF} $\widehat Y_{\rep s}^{(n_Y)}(T)$ with modular weight $n_Y\in\mathds{N}$ transform under $\gamma\in\SL{2,\Z{}}_T$ according to
\begin{equation}
 \widehat Y_{\rep s}^{(n_Y)}(T) ~\xmapsto{~\gamma~}~  \widehat Y_{\rep s}^{(n_Y)}\left(\frac{aT+b}{cT+d}\right)
                                ~=~ (c\,T + d)^{n_Y} \rho_{\rep s}(\gamma)\, \widehat Y_{\rep s}^{(n_Y)}(T)\,,
\end{equation}
where $\rho_{\rep s}(\gamma)$ is an $s$-dimensional representation of $T'$. These modular forms can be built from tensor products of
the lowest weight ($n_Y=1$) \ac{VVMF}~\cite{Liu:2019khw}
\begin{equation}
 \widehat Y^{(1)}_{\rep2''}~:=~  \begin{pmatrix} -3\sqrt{2}\frac{\eta^3(3T)}{\eta(T)} \\ 3\frac{\eta^3(3T)}{\eta(T)}+\frac{\eta^3(T/3)}{\eta(T)} \end{pmatrix}\,,
\end{equation}
where $\eta(T)$ is the so-called Dedekind $\eta$-function.
Using the transformations under the modular generators $\mathrm S,\mathrm T$, it is easy to confirm that $\widehat Y^{(1)}_{\rep2''}(T)$ transforms
as a doublet $\rep2''$ of $T'$~\cite{Nilles:2020kgo}.
Modular forms of higher weight can be constructed by the tensor product of this lowest weight modular form doublet.
For example, we know that in general  $\rep2''\otimes\rep2''=\rep3\oplus\rep1'$ but $\rep1'$ vanishes here for identical doublets
in $T'$. Hence, $\widehat Y^{(1)}_{\rep2''}\otimes\widehat Y^{(1)}_{\rep2''}$ delivers a \ac{VVMF} $\widehat Y^{(2)}_{\rep3}$.

As we mentioned earlier, the superpotential must have modular weight $-1$, this implies that trilinear couplings built by
$\Phi_{n}$ multiplets require modular forms of weight $n_Y=-1-3n$. That is, $\Phi_{-\nicefrac23}^3$ would require the
coupling given by $\widehat Y^{(1)}_{\rep{s}}$ while $\Phi_{-\nicefrac53}^3$ needs $\widehat Y^{(4)}_{\rep{s}}$. We observe that the former case
includes $T'$ doublets while the latter does not.

Finally, let us point out that $T'$ is the double cover of $A_4$. Therefore, the modular forms of $T'$ include those of $A_4$.
The \acp{VVMF} $\widehat Y^{(n_Y)}_{\rep{s}}$ with even $n_Y$  build irreducible representations of $A_4$.

%%%%%%%%%%%%%%%%%%%%%%%%%%%%%%%%%%%%%%%%%%%%%%%%%%%%%%%%%%%%%%%%%%%%%%%%%%
%  Trilinear couplings
%%%%%%%%%%%%%%%%%%%%%%%%%%%%%%%%%%%%%%%%%%%%%%%%%%%%%%%%%%%%%%%%%%%%%%%%%%

\section{Results and insights from trilinear couplings}
\label{sec:trilinear}

To illustrate the power of the eclectic symmetry, we can use the allowed trilinear couplings of twisted fields in the $\mathds T^2/\Z3$ orbifold computed earlier~\cite{Nilles:2020kgo}. The relevant fields are $\Phi^i_{-\nicefrac23}=(X_i,Y_i,Z_i)^\mathrm{T}$ and   $\Phi^i_{-\nicefrac53}=(\tilde X_i,\tilde Y_i,\tilde Z_i)^\mathrm{T}$, where the index $i$ labels different field multiplets. They both transform as a $\rep1\oplus\rep2'$  representation of the modular group $T'$ but have different modular weight. The difference of the couplings for $\Phi^i_{-\nicefrac23}$ and $\Phi^i_{-\nicefrac53}$ will therefore be due to the appearance of the hybrid $\Z2^\mathrm{hybrid}$ symmetry connected to the $\mathrm{S}^2=-\Id$ transformation in \SL{2,\Z{}}. This is not an element of \PSL{2,\Z{}} and does not transform the modulus. Superficially, it can thus be understood as a traditional flavor symmetry, although it is intrinsically modular. This transformation $\mathrm{C}=\mathrm{S}^2$ is correlated with the modular weights as shown in \Cref{eq:C=S2rep}. Under $\Delta(54)$, $\Phi^i_{-\nicefrac23}$ transforms as an irreducible $\rep3_2$ representation while $\Phi^i_{-\nicefrac53}$ transforms as a $\rep3_1$. The difference comes again from the hybrid $\Z2^\mathrm{hybrid}$ that extends $\Delta(27)$ to $\Delta(54)$. In the following, we shall discuss the trilinear couplings from two different angles. First, in \Cref{subsec:RestrictionsModular}, we start by imposing the modular flavor symmetry $T'$.
Then, in \Cref{subsec:RestrictionsTraditional}, we study the couplings based on the $\Delta(54)$ traditional flavor symmetry.

Our aim is to check the validity of Rule 4 (see \Cref{sec:rule4}), which predicts a different behavior for the trilinear couplings of $(\Phi_{-\nicefrac23})^3$ and $(\Phi_{-\nicefrac53})^3$. This is because  $\Phi_{-\nicefrac53}$ contains an oscillator mode while $\Phi_{-\nicefrac23}$ does not. Rule 4 would then by definition require the absence of the trilinear couplings $\tilde X^3$, $\tilde Y^3$ and $\tilde Z^3$ while $X^3$, $Y^3$ and $Z^3$ should be allowed.

%%%%%%%%%%%%%%%%%%%%%%%%%%%%%%
\subsection{Restrictions from modular symmetry}
\label{subsec:RestrictionsModular}

Possible trilinear terms of twisted fields in the superpotential are given by
\begin{equation}
\mathcal{W} ~\supset~ \widehat Y^{(1)}(T)\,\Phi_{-\nicefrac23}^{1}\Phi_{-\nicefrac23}^{2}\Phi_{-\nicefrac23}^{3} +
          \widehat Y^{(4)}(T)\,\Phi_{-\nicefrac53}^{1}\Phi_{-\nicefrac53}^{2}\Phi_{-\nicefrac53}^{3}\;,
\end{equation}
with modular forms $\widehat Y^{(n_Y)}$ of weight $n_Y=1$ and
$n_Y=4$. As mentioned earlier in~\Cref{subsec:TprimerForms},
the $T'$ transformation of the modular forms can be deduced through
products of the ``fundamental'' modular form $\widehat Y^{(1)}_{\rep2''}(T)$,
which transforms as a $\rep2''$ of $T'$. $\widehat Y^{(4)}(T)$ as the
fourth power of $\rep2''$ then transforms as a $\rep1\oplus\rep1'\oplus\rep3$-dimensional
representation of $T'$. These couplings have been worked out
in Ref.~\cite{Nilles:2020kgo} and turn out to be 
\begin{subequations}
\label{eq:WtermsNoOscillators}
\begin{eqnarray}
\mathcal{W}_1 & = & \frac{1}{4} \left(\widehat{Y}_2(T) (4\, X_1\, X_2\, X_3 + (Y_1+Z_1) (Y_2+Z_2) (Y_3+Z_3))\right. \\
              &   & \left.-\sqrt{2} \widehat{Y}_1(T) \left((Y_1+Z_1) (Y_2+Z_2)X_3 + ((Y_1+Z_1)X_2 + X_1 (Y_2+Z_2)) (Y_3+Z_3)\right)\right)\;, \nonumber \\
\mathcal{W}_2 & = & \frac{1}{4} \left(\sqrt{2} \widehat{Y}_1(T) X_1 + \widehat{Y}_2(T) (Y_1+Z_1)\right) (Y_2-Z_2) (Y_3-Z_3)\;,\\
\mathcal{W}_3 & = & \frac{1}{4} (Y_1-Z_1) \left(\sqrt{2} \widehat{Y}_1(T) X_2 + \widehat{Y}_2(T) (Y_2+Z_2)\right) (Y_3-Z_3)\;,\\
\mathcal{W}_4 & = & \frac{1}{4} (Y_1-Z_1) (Y_2-Z_2) \left(\sqrt{2} \widehat{Y}_1(T) X_3 + \widehat{Y}_2(T) (Y_3+Z_3)\right)\;,
\end{eqnarray}
\end{subequations}
for the twisted matter fields $\Phi^i_{-\nicefrac23}=(X_i,Y_i,Z_i)^\mathrm{T}$ and
\begin{subequations}
\label{eq:WtermsOscillators}
\begin{eqnarray}
\widetilde{\mathcal{W}}_1 & = & \tfrac{1}{2\sqrt2} \widehat{Y}_{\rep1'}^{(4)}(T) \left(\tilde X_2 (\tilde Y_1 + \tilde Z_1) - \tilde X_1 (\tilde Y_2 + \tilde Z_2) \right) (\tilde Y_3 - \tilde Z_3)\,,\\
\widetilde{\mathcal{W}}_2 & = & \tfrac{1}{2\sqrt2} \widehat{Y}_{\rep1'}^{(4)}(T) \left(\tilde X_3 (\tilde Y_1 + \tilde Z_1) - \tilde X_1 (\tilde Y_3 + \tilde Z_3) \right) (\tilde Y_2 - \tilde Z_2)\,,\\
\widetilde{\mathcal{W}}_3 & = & \tfrac{1}{2\sqrt2} \widehat{Y}_{\rep1'}^{(4)}(T) \left(\tilde X_3 (\tilde Y_2 + \tilde Z_2) - \tilde X_2 (\tilde Y_3 + \tilde Z_3) \right) (\tilde Y_1 - \tilde Z_1)\,,\\
\widetilde{\mathcal{W}}_4 & = & \tfrac{1}{2\sqrt2} \widehat{Y}_{\rep1}^{(4)}(T) \,(\tilde Y_1 - \tilde Z_1) (\tilde Y_2 - \tilde Z_2) (\tilde Y_3 - \tilde Z_3)\,,\\
\widetilde{\mathcal{W}}_5 & = & \tfrac{1}{2\sqrt2} (\tilde Y_3 - \tilde Z_3) \bigg[ \tilde X_2 \left(2\,\widehat Y_{\rep3,3}^{(4)}(T) \tilde X_1 + \widehat Y_{\rep3,2}^{(4)}(T)(\tilde Y_1 + \tilde Z_1) \right) \\
              &   & \hspace{2.5cm} +\, (\tilde Y_2 + \tilde Z_2) \left(\widehat Y_{\rep3,2}^{(4)}(T) \tilde X_1 + \widehat Y_{\rep3,1}^{(4)}(T) (\tilde Y_1 + \tilde Z_1) \right)\bigg]\,, \nonumber\\
\widetilde{\mathcal{W}}_6 & = & \tfrac{1}{2\sqrt2} (\tilde Y_2 - \tilde Z_2) \bigg[ \tilde X_3 \left(2\,\widehat Y_{\rep3,3}^{(4)}(T) \tilde X_1 + \widehat Y_{\rep3,2}^{(4)}(T)(\tilde Y_1 + \tilde Z_1) \right) \\
              &   & \hspace{2.5cm} +\, (\tilde Y_3 + \tilde Z_3) \left(\widehat Y_{\rep3,2}^{(4)}(T) \tilde X_1 + \widehat Y_{\rep3,1}^{(4)}(T) (\tilde Y_1 + \tilde Z_1) \right)\bigg]\,, \nonumber\\
\widetilde{\mathcal{W}}_7 & = & \tfrac{1}{2\sqrt2} (\tilde Y_1 - \tilde Z_1) \bigg[ \tilde X_3 \left(2\,\widehat Y_{\rep3,3}^{(4)}(T) \tilde X_2 + \widehat Y_{\rep3,2}^{(4)}(T)(\tilde Y_2 + \tilde Z_2) \right) \\
              &   & \hspace{2.5cm} +\, (\tilde Y_3 + \tilde Z_3) \left(\widehat Y_{\rep3,2}^{(4)}(T) \tilde X_2 + \widehat Y_{\rep3,1}^{(4)}(T) (\tilde Y_2 + \tilde Z_2) \right)\bigg]\,, \nonumber
\end{eqnarray}
\end{subequations}
for the twisted fields $\Phi^i_{-\nicefrac53}=(\tilde X_i,\tilde Y_i,\tilde Z_i)^\mathrm{T}$.
These expressions differ significantly even though both fields
$\Phi^i_{-\nicefrac23}$ and $\Phi^i_{-\nicefrac53}$ are in the same
$\rep1\oplus\rep2'$-representation of $T'$. The difference is due to the
different modular weights and thus the action of the hybrid
$\Z2^\mathrm{hybrid}$ symmetry (cf.\ \Cref{eq:C=S2rep}).

What can we learn from these results regarding the validity of Rule 4?
For the trilinear couplings of $\Phi^i_{-\nicefrac23}$, we see that the self-couplings $X^3$, $Y^3$ and $Z^3$ are all allowed.
This is compatible with Rule 4. The result for the self-couplings of $\Phi^i_{-\nicefrac53}$ is less transparent.
A closer inspection of formula~\eqref{eq:WtermsOscillators} reveals the fact that $\tilde X^3$ is forbidden
while the couplings $\tilde Y^3$ and $\tilde Z^3$ are still allowed.
What is the origin of this difference?
A look at the $T'$ representations~\eqref{eq:2prime+1} gives the answer. $\tilde X$ is exclusively a member of the doublet,
while $\tilde Y$ and $\tilde Z$ appear both in the doublet and singlet representations of $T'$.
Still, we see that the result does not respect Rule 4 as $\tilde Y^3$ and $\tilde Z^3$ are still allowed.

But this is not yet the end of the story as we still have traditional flavor symmetries
at our disposal. These include  a $\Z3^\mathrm{rot}$ symmetry
that rotates the twisted fields as discussed in
Figure~2 of~\cite{Baur:2019kwi}.
It can be understood as an outer automorphism of the
space group of the orbifolded \Z3 lattice. If we apply these
restrictions on the couplings given in \Cref{eq:WtermsNoOscillators,eq:WtermsOscillators},
we obtain
\begin{align}
\label{eq:Z3superpotential}
\mathcal{W}(T,X_i,Y_i,Z_i) & ~\supset~ c^{(1)}\, \Big[\widehat{Y}_2(T) \big( X_1\,X_2\,X_3 + Y_1\,Y_2\,Y_3 + Z_1\,Z_2\,Z_3\big)\nonumber \\
                          &  \hspace{-20mm} - \frac{\widehat{Y}_1(T)}{\sqrt{2}} \big( X_1\,Y_2\,Z_3 + X_1\,Y_3\,Z_2 + X_2\,Y_1\,Z_3
                                    +\, X_3\,Y_1\,Z_2 + X_2\,Y_3\,Z_1 + X_3\,Y_2\,Z_1\big)\Big]\;,
\end{align}
and
\begin{equation}
\label{eq:Z3superpotential-withoscillators}
\widetilde{\mathcal{W}}(T,\tilde X_i,\tilde Y_i,\tilde Z_i) \supset c^{(4)}\, \widehat Y^{(4)}_{\rep1'}(T)
               \Big(  \tilde X_1\,\tilde Y_3\,\tilde Z_2 - \tilde X_1\,\tilde Y_2\,\tilde Z_3 + \tilde X_2\,\tilde Y_1\,\tilde Z_3
  -\tilde X_2\,\tilde Y_3\,\tilde Z_1 + \tilde X_3\,\tilde Y_2\,\tilde Z_1 - \tilde X_3\,\tilde Y_1\,\tilde Z_2\Big).
\end{equation}
respectively.\footnote{Note that this $\Z3^\mathrm{rot}$ symmetry is sufficient to obtain this result, and the $\Z3^\mathrm{B}$ symmetry is automatically satisfied. We do not need the full traditional flavor symmetry $\Delta(54)$ or $\Delta(27)$ for this result}. \Cref{eq:Z3superpotential,eq:Z3superpotential-withoscillators} are now compatible with Rule 4. The difference between these equations is mainly a consequence of the hybrid $\Z2^\mathrm{hybrid}$ symmetry $\mathrm{C}=\mathrm{S}^2$. In combination with the rotational $\Z3^\mathrm{rot}$ symmetry, it appears to provide the origin of Rule 4 in the case of trilinear couplings. A geometrical explanation of both, the hybrid $\Z2^\mathrm{hybrid}$ and
the rotational $\Z3^\mathrm{rot}$ symmetries, has been given in \cite[Figure~2]{Baur:2019kwi},
where the \Z2 symmetry appears as a 180-degree rotation of the \Z3 lattice. There, it was shown that these two symmetries combine
to the non-Abelian $R$-symmetry $S_3^R$. Hence, this $S_3^R$ is crucial for an explanation of Rule 4.

%%%%%%%%%%%%%%%%%%%%%%%%%%%%%%
\subsection{Restrictions from the traditional flavor symmetry}
\label{subsec:RestrictionsTraditional}

Let us now analyze the restrictions from the traditional flavor symmetry $\Delta(27)$ and
$\Delta(54)$ in more detail. Both twisted fields $\Phi_{-\nicefrac23}$
and $\Phi_{-\nicefrac53}$ are in the irreducible triplet representation
of $\Delta(27)$ and should have identical trilinear couplings
in the superpotential. Within $\Delta(54)$, however,
$\Phi_{-\nicefrac23}$ and $\Phi_{-\nicefrac53}$ transform as different
3-dimensional representations, $\rep3_2$ and $\rep3_1$, respectively.
This leads to a decisive difference in the trilinear Yukawa
couplings. The superpotential transforms as a $\rep1'$ representation
of $\Delta(54)$. We thus have to consider possible $\rep1'$
representations in the the product $\rep3_i\otimes\rep3_i\otimes\rep3_i$ for $i=1,2$.
As shown explicitly in~\Cref{app:Delta54}, tensor products
for the triplets are given by
\begin{equation}
\rep3_1\otimes \rep3_1 ~=~ \rep3_2\otimes \rep3_2 ~=~ \crep3_1 \oplus \crep3_1 \oplus \crep3_2\,.
\end{equation}
Nontrivial singlets are contained in the product of
$\crep3_1 \otimes \rep3_2$ and $\crep3_2 \otimes \rep3_1$. For the
product $\left(\Phi_{-\nicefrac23}\right)^3=\rep3_2\otimes\rep3_2\otimes\rep3_2$, we thus have two types of
couplings $\crep3_1\otimes \rep3_2$, while for
$\left(\Phi_{-\nicefrac53}\right)^3=\rep3_1 \otimes\rep3_1\otimes\rep3_1$ there is only one, namely $\crep3_2\otimes \rep3_1$.
In an explicit calculation using the tensor products of \Cref{app:Delta54}, we
find that couplings of three $\rep3_1\otimes\rep3_1\otimes\rep3_1$ at the same fixed points
are forbidden while those of the three $\rep3_2\otimes\rep3_2\otimes\rep3_2$ are allowed.
Crucial for this selection rule is again the hybrid $\Z2^\mathrm{hybrid}$ symmetry
that, in this case,  enhances $\Delta(27)$ to $\Delta(54)$
(which in turn can be identified with the symmetry $\mathrm{C}=\mathrm{S}^2$).
We thus see, that the traditional flavor symmetry forbids trilinear couplings for
fields $\tilde X, \tilde Y, \tilde Z$
at them same fixed point. It does not give, however, the
full structure of formulae \Cref{eq:Z3superpotential,eq:Z3superpotential-withoscillators},
which contain more information about  the moduli-dependence of the
Yukawa-couplings (which is not restricted by $\Delta(54)$).

As we have mentioned before, a geometric origin of this \Z2
can also be found in the \Z2-outer automorphism of the space
group selection rule of the $\mathds{T}^2/\Z3$ lattice (see Figure~2 of~\cite{Baur:2019kwi}).
This reflects the fact that the \Z3 lattice
has the same symmetries as the \Z6 lattice.
At the technical level, this is a consequence of the fact that the correlation functions involve
a sum over sublattices, and the sublattice of the local coupling  has an additional $\Z2$
symmetry compared to the sublattice relevant for the nonlocal coupling.

%%%%%%%%%%%%%%%%%%%%%%%%%%%%%%
\subsection{Lessons from trilinear couplings}

Explicit calculations show that Rule 4 (in trilinear couplings) is the consequence of symmetries. A crucial role is played by a hybrid $\Z2^\mathrm{hybrid}$ and a rotational $\Z3^\mathrm{rot}$ symmetry that combine to the group $S_3^R$. Rule 4 can be reproduced in two dual ways: either by starting from the modular symmetry $T'$ and invoking additionally the $\Z3^\mathrm{rot}$ symmetry, or by considering the full $\Delta (54)$ traditional symmetry. In both cases, the hybrid $\Z2^\mathrm{hybrid}$ and the rotational $\Z3^\mathrm{rot}$ symmetries play a central role. In this discussion, the particular properties of the fields $X_i$ and $\tilde X_i$ play an important role.

%%%%%%%%%%%%%%%%%%%%%%%%%%%%%%%%%%%%%%%%%%%%%%%%%%%%%%%%%%%%%%%%%%%%%%%%%%
%  Higher order
%%%%%%%%%%%%%%%%%%%%%%%%%%%%%%%%%%%%%%%%%%%%%%%%%%%%%%%%%%%%%%%%%%%%%%%%%%

\section{General proof for higher-order couplings}
\label{sec:higherOrder}

We now proceed to study Rule 4 for general $n$-point couplings in \Z3 orbifolds.
Let us consider higher order couplings $(\Phi_{-\nicefrac23})^n$ and $(\Phi_{-\nicefrac53})^n$, $n>3$,
for the fields $\Phi^i_{-\nicefrac{2}{3}}=(X_i,Y_i,Z_i)^{\mathrm{T}}$ and $\Phi^i_{-\nicefrac{5}{3}}=(\tilde X_i,\tilde Y_i,\tilde Z_i)^{\mathrm{T}}$.
$(\Phi_{-\nicefrac23})^n$ leads to $\Sigma=n-3$ and, as stated in \Cref{sec:rule4}, Rule 4 requires $\Sigma = 6\ell$ with integer $\ell$.
Thus, in this case, $n=3+6\ell$ for the allowed {\it local} couplings $X^n, Y^n, Z^n$ at the same fixed point.
On the other hand, $\Phi_{-\nicefrac53}$ contains one oscillator excitation; hence, $(\Phi_{-\nicefrac53})^n$ leads to
$\Sigma=n-3+n=2n-3$, which should be $0\bmod 6$ for allowed local couplings.
As $2n-3$ is always an odd integer, there is no solution for $2n-3=6\ell$ ($\ell$ integer).

In order to arrive at a possible symmetry explanation of this outcome, we now
follow the procedure applied for trilinear couplings, i.e.\ we start by relating it to the modular properties of the
theory, and then we consider its possible origin within the $\Delta(54)$ traditional flavor symmetry.

%%%%%%%%%%%%%%%%%%%%%%%%%%%%%%
\subsection{Constraints from modular symmetry}

The total modular weights of the couplings $(\Phi_{-\nicefrac23})^n$ and $(\Phi_{-\nicefrac53})^n$
are $-\nicefrac{2n}{3}$ and $-\nicefrac{5n}{3}$, respectively. Moreover, the modular weight of the superpotential must be $-1$.
This implies that the modular weights of the corresponding couplings strengths, which are \acp{VVMF}, should be
$n_Y = -1 + \nicefrac{2n}{3}$ and $n_Y = -1 + \nicefrac{5n}{3}$.
Since $n_Y\in\Z{}$ for the \acp{VVMF} of $\Gamma_3'\cong T'$, this results in $n = 3 k$ for integer $k$.
We note in addition that this constraint also follows from the space and point-group selection rules (see e.g.~\cite{Kobayashi:2011cw}).

To simplify the discussion, let us restrict ourselves to the local couplings of the twisted fields, as they are of
relevance for Rule 4. This leads to
\begin{equation}
\label{eq:WlocalY}
\mathcal{W}_\mathrm{local} = \sum_{k_1, k_2>0} \widehat{Y}^{(2k_1-1)}_{\mathrm{local}}(T)\, (X^{3k_1} + Y^{3k_1}+Z^{3k_1}) +
\tilde{Y}^{(5k_2-1)}_{\mathrm{local}}(T)\, (\tilde{X}^{3k_2}+\tilde{Y}^{3k_2}+\tilde{Z}^{3k_2}) \;.
\end{equation}
As we have seen before, the fields $X$ and $\tilde X$ play a special role as they belong exclusively to
the doublet representation $\rep2'$ of $T'$. We thus consider
\begin{equation}
\label{eq:WlocalX}
\mathcal{W}_\mathrm{local} \supset \sum_{k_1, k_2>0} \widehat{Y}^{(2k_1-1)}_{\mathrm{local}} X^{3 k_1} + \widehat{Y}^{(5k_2-1)}_{\mathrm{local}} \tilde{X}^{3 k_2}\;.
\end{equation}
The representations of the modular forms  $\widehat Y^{(n_Y)}$ of $T'$ (cf.\ discussion on~\cite[Section 3.2]{Liu:2019khw}) fall into two classes with respect to the $\Z2^\mathrm{hybrid}$.
For even modular weights, they correspond to the $A_4$ representations $\rep1,\rep1',\rep1'',\rep3$, which we will refer to as corresponding to the ``even class'' of $\Z2^\mathrm{hybrid}$. For odd modular weights,
they belong to the doublet representations $\rep2,\rep2',\rep2''$ of $T'$, the double cover of $A_4$, which we will refer to as corresponding to the ``odd class'' of $\Z2^\mathrm{hybrid}$. The products of
fields and modular forms in the superpotential have to combine to the trivial singlet of $T'$, corresponding
to the even class with respect to the hybrid $\Z2^\mathrm{hybrid}$.

$X^{3k_1}$ has weight $-2k_1$. As the superpotential has weight $-1$, the weight of $\widehat Y^{(n_Y)}$
has to be $n_Y=2k_1-1$ which is always in the odd class (only doublets). The class of $X^{3k_1}$ is
odd (even)  if $k_1$ is odd (even). Thus, for odd $k_1$ the couplings are allowed while for even $k_1$
they are forbidden. For the local couplings $X^n$,s this reproduces the $n=3\bmod 6$ selection rule
of Rule 4.

$\tilde X^{3k_2}$ has weight $-5k_2$. For $\widehat Y^{(n_Y)}$ we thus obtain weight $n_Y=5k_2-1$. The
product $\widehat Y^{(5k_2-1)} \tilde X^{3k_2}$ has modular weight $8k_2-1$. This is always in the
odd class of $\Z2^\mathrm{hybrid}$ and therefore not allowed as a term of the superpotential,
in agreement with Rule 4.

For the other twisted fields $Y, Z, \tilde Y, \tilde Z$ this argumentation does not hold as they
are not exclusively members of the $\rep2'$ representation of $T'$ (but also appear in the singlet representation).
To complete the proof of Rule 4 we thus again need the rotational $\Z3^\mathrm{rot}$ traditional flavor symmetry.
Hybrid $\Z2^\mathrm{hybrid}$ and rotational $\Z3^\mathrm{rot}$ are the crucial symmetries to explain Rule 4.

%%%%%%%%%%%%%%%%%%%%%%%%%%%%%%
\subsection{Constraints from traditional flavor symmetry}

Let us again concentrate on local couplings. The fields $\Phi_{-\nicefrac23}$ and $\Phi_{-\nicefrac53}$ are both triplets of $\Delta(27)$
and differ only in their transformation properties under the hybrid $\Z2^\mathrm{hybrid}$ generated by $\mathrm C=\mathrm S^2$. If we impose $\Delta(27)$ we obtain
 \begin{equation}
\label{eq:ABinvarianceW}
\mathcal{W}_\mathrm{local} =\sum_{k_1,k_2>0} a_{k_1} (X^{3k_1} + Y^{3k_1}+Z^{3k_1}) + b_{k_2} (\widetilde{X}^{3k_2}+\widetilde{Y}^{3k_2}+\widetilde{Z}^{3k_2}) \;,
\end{equation}
with general coefficients $a_{k_1}, b_{k_2}$. $\Delta(27)$ includes the rotational $\Z3^\mathrm{rot}$ symmetry
and this manifests itself in the fact that the coefficients for the terms with $X,Y,Z$ and
$\tilde X, \tilde Y, \tilde Z$ are universal. We remark that the powers $3k_1$ and $3k_2$ here arise from the $\Z3^{\mathrm{B}}$ (or $\Z3^{\mathrm{A^2B^2AB}}$) factor of $\Delta(27)$.
Under the generator $\mathrm C$ of the hybrid $\Z2^\mathrm{hybrid}$ the
superpotential changes sign. In addition, the field $\Phi_{-\nicefrac23}$ changes sign while $\Phi_{-\nicefrac53}$
does not. It follows that
\begin{align}
\mathcal{W}_\mathrm{local} \xmapsto{~\mathrm{C}~} &\sum_{k_1, k_2>0} (-1)^{3k_1}a_{k_1} (X^{3k_1} +Z^{3k_1} +Y^{3k_1}) + b_{k_2} (\tilde{X}^{3k_2}+\tilde{Z}^{3k_2}+\tilde{Y}^{3k_2}) \\
&\stackrel!= -\mathcal{W}_\mathrm{local} = \sum_{k_1, k_2>0} - a_{k_1} (X^{3k_1} +Y^{3k_1}+Z^{3k_1} )- b_{k_2} (\tilde{X}^{3k_2}+\tilde{Y}^{3k_2}+\tilde{Z}^{3k_2})  \;.
\end{align}
This implies that the $b$ coefficients have to vanish and that $3k_1$ has to be odd. This
reproduces Rule 4. Again the hybrid $\Z2^\mathrm{hybrid}$ plays a crucial role. 
The rotational $\Z3^\mathrm{rot}$ (as discussed in the
last section) is relevant again, as it is included in $\Delta(27)$. Hybrid $\Z2^\mathrm{hybrid}$ and rotational $\Z3^\mathrm{rot}$ are at the heart of this solution. This is of truly eclectic nature as an interplay of the traditional \Z3 symmetry and the modular \Z2 symmetry. They combine to the non-Abelian group $S_3$. As the superpotential transforms nontrivially under \Z2 this is an $R$-symmetry $S^R_3$. This explains why this solution has been missed in the earlier discussion~\cite{Font:1988nc}, as there only Abelian $R$-symmetries had been considered.

%%%%%%%%%%%%%%%%%%%%%%%%%%%%%%%%%%%%%%%%%%%%%%%%%%%%%%%%%%%%%%%%%%%%%%%%%%
%  Consequences
%%%%%%%%%%%%%%%%%%%%%%%%%%%%%%%%%%%%%%%%%%%%%%%%%%%%%%%%%%%%%%%%%%%%%%%%%%
\section{Some immediate consequences}
\label{sec:consequences}

As long as Rule 4 was considered as a ``stringy miracle'', its range of validity was not clear.
In many applications, it was thus assumed that this Rule 4 would also hold under more 
general circumstances as, for example, in the presence of background fields such as Wilson lines.
Now that we know that Rule 4 has its origin in a non-Abelian $R$-symmetry, we can analyze more
general situations.

Properties of Wilson lines have been first discussed in Ref.~\cite{Ibanez:1986tp}. There, it was shown that such
background fields do not only break gauge symmetries (in the heterotic orbifold picture), but also lift
the degeneracy of the fixed points. This second property was important for the construction of
models with 3 families of quarks and leptons. In our previous discussion the degeneracy of the fixed points was the result of the rotational $\Z3^\mathrm{rot}$ symmetry. Thus, Wilson lines break this symmetry (and with it $\Delta(54)$). An analysis of the modular flavor symmetry reveals the fact that also $\SL{2,\Z{}}$ is generically broken by Wilson lines~\cite{Bailin:1993ri}. This would then imply that the finite modular flavor symmetry $T'$ is broken as well. As both $\Z3^\mathrm{rot}$ and $T'$ have been crucial for the proof of Rule 4, we would then have to worry that such a rule might not hold in the presence of Wilson lines. It seems that the full beauty of the eclectic flavor symmetry needs the consideration of models like those constructed in Refs.~\cite{Carballo-Perez:2016ooy,Baur:2022hma}, where one of the three two-tori $\mathds T^2$ (in six dimensional) compactified space does not feel the Wilson line background fields.

%%%%%%%%%%%%%%%%%%%%%%%%%%%%%%%%%%%%%%%%%%%%%%%%%%%%%%%%%%%%%%%%%%%%%%%%%%
%  Conclusions
%%%%%%%%%%%%%%%%%%%%%%%%%%%%%%%%%%%%%%%%%%%%%%%%%%%%%%%%%%%%%%%%%%%%%%%%%%

\section{Summary and Outlook}
\label{sec:conclusions}

We have seen that the presence of duality transformations in string theory can manifest itself in
a variety  of restrictions in the low-energy effective action. Some of them can be understood through
the appearance of certain modular symmetries derived from the modular group $\SL{2,\Z{}}$.
Among these are:
\begin{itemize} 
\item The appearance of a (nonlinearly realized) discrete modular symmetry.

\item It is accompanied by a discrete linearly realized traditional flavor symmetry.

\item The discrete modular group can be understood in a bottom-up approach 
through the outer automorphisms of the traditional flavor group.

\item Both combine to form an eclectic flavor group.

\item At some specific points in moduli space there are enhancements of the linearly
realized discrete symmetry.
\end{itemize}

Apart from these somewhat obvious restrictions, we are led to some surprises. The full power of the modular symmetry is encoded in the modular forms that appear as coefficients in  the Yukawa couplings. These constraints are indirect and appear somewhat intransparent from a general field theoretic point of view. Technically, the rules can be easily implemented through a construction of the modular forms as products of the basic modular forms of lowest weight~\cite[around Equation (5.8)]{Liu:2021gwa}. In the \Z3 case, the lowest modular form at weight one builds a $\rep2''$ representation of $T'$. Its products lead to a $\rep3$ at weight two, but the singlet is missing. The absence of certain representations continues at higher weight.
The direct consequences of these observations are not yet fully understood and need further investigations.

In the present paper, we have discussed in detail the explanation of an enigmatic selection rule in string compactifications, Rule 4, in the $\mathds{T}^2/\Z3$ orbifold through modular and conventional flavor symmetries. The important lesson is not just
the explanation of this selection rule, but the mechanism how traditional and modular
symmetries combine to lead to this result. In particular, we showed that Rule 4 can be explained in two different ways through a non-Abelian $R$-symmetry. In the first case, Rule 4 can be explained through the $T'$ modular group (including the $\Z2^{\mathrm{hybrid}}$ symmetry), the point group $\Z{3}^{\mathrm{(PG)}}$ and the traditional flavor symmetry $\Z3^{\mathrm{rot}}$. In the second case, Rule 4 is explained solely by the traditional flavor group $\Delta(54)$. In both cases, the key component to explain Rule 4 is the non-Abelian $R$-symmetry given by $S_3^R \cong \Z3^{\mathrm{rot}} \rtimes \Z2^{\mathrm{hybrid}}$. This explains why Rule 4 could not be obtained from conventional symmetries in~\cite{Font:1988nc}, which did not considered non-Abelian $R$-symmetries.

This seems to be just the tip of the iceberg though. Many questions are still open.
The modular forms ``know'' that the nonlinearly realized
modular symmetries get enhanced at some specific points in moduli space.\footnote{The
explicit form of the superpotential at these specific points can be found in e.g.\ Table 4
and Equation (3.59) of Ref.~\cite{Li:2025bsr}.} What is the role
of these symmetries in the interior of moduli space? Is there something specific in the 
nonlinear realization of modular symmetries that is not shared by nonlinearly realized
symmetries that appear from a spontaneous breakdown of a linearly realized symmetry
via a Higgs-mechanism?

One of the important aspects of the system seems to be the role of outer automorphisms
of various symmetry groups.
In the orbifold construction, outer automorphisms of the lattice lead to the space-group
selection rule and outer automorphisms of the Narain space group complete the eclectic flavor group.
From a bottom-up approach the discrete modular flavor group is connected to the outer automorphisms
of the traditional discrete flavor group. It should be further analyzed how these properties
can influence the presence or absence of couplings in the low-energy effective theory.

\section*{Acknowledgments}
We acknowledge the passionate discussions with Michael Ratz at various stages of this project,
where we came to realize that different interpretations of the presented observations can be drawn.
The work by VKP and XGL was supported in part by the U.S. National Science Foundation under
Grant PHY-2210283. SRS is partly supported by UNAM-PAPIIT IN113223 and IN117226, and Marcos Moshinsky Foundation.

%%%%%%%%%%%%%%%%%%%%%%%%%%%%%%%%%%%%%%%%%%%%%%%%%%%%%%%%%%%%%%%%%%%%%%%%%%
%  Appendix
%%%%%%%%%%%%%%%%%%%%%%%%%%%%%%%%%%%%%%%%%%%%%%%%%%%%%%%%%%%%%%%%%%%%%%%%%%
\appendix

\section{Some group theoretical details}
\label{app:Details}

\subsection[Group T' = (24,3)]{\boldmath Group $T'\cong[24,3]$ \unboldmath}
\label{app:Tprime}

The finite modular group $\Gamma'_3\cong T'\cong Q_8\rtimes\Z3\cong[24,3]$ has 24 elements,
which can be generated by the generators $\mathrm S, \mathrm T$, satisfying the presentation
\begin{equation}
 T' ~=~ \left\langle \mathrm S, \mathrm T ~|~ \mathrm S^4 = (\mathrm S\mathrm T)^3 = \mathrm T^3 = \mathrm S^2 \mathrm T\mathrm S^{-2}\mathrm T^{-1} = \Id  \right\rangle\;.
\end{equation}
Its irreducible representations are a triplet $\rep3$, three doublets $\rep2,\rep2',\rep2''$ and three singlets $\rep1,\rep1',\rep1''$.
The corresponding representation matrices are shown in Table~\ref{tab:irrepTprime} below.
\begin{table}[th!]
\centering
\begin{tabular}{ccc}
\toprule 
Irrep & $\rho_{\mathbf{r}}(\mathrm{S})$ & $\rho_{\mathbf{r}}(\mathrm{T})$ \\
\midrule $\rep1$ & 1 & 1 \\
$\rep{1'}$ & 1 & $\omega$ \\
$\rep1^{\prime \prime}$ & 1 & $\omega^2$ \\ \midrule
$\rep2$ & $\frac{\mathrm{i}}{\sqrt{3}}\left(\begin{array}{cc}1 & \sqrt{2} \\
\sqrt{2} & -1\end{array}\right)$ & $\omega\left(\begin{array}{cc}1 & 0 \\
0 & \omega\end{array}\right)$ \\
$\rep2^{\prime}$ & $\frac{\mathrm{i}}{\sqrt{3}}\left(\begin{array}{cc}1 & \sqrt{2} \\
\sqrt{2} & -1\end{array}\right)$ & $\omega^2\left(\begin{array}{ll}1 & 0 \\
0 & \omega\end{array}\right)$ \\
$\rep2^{\prime\prime}$ & $\frac{\mathrm{i}}{\sqrt{3}}\left(\begin{array}{cc}1 & \sqrt{2} \\
\sqrt{2} & -1\end{array}\right)$ & $\left(\begin{array}{cc}1 & 0 \\
0 & \omega\end{array}\right)$ \\ \midrule
$\rep3$ & $\frac{1}{3}\left(\begin{array}{ccc}-1 & 2 & 2 \\
2 & -1 & 2 \\
2 & 2 & -1\end{array}\right)$ &
$\left(\begin{array}{ccc}1 & 0 & 0 \\
0 & \omega & 0 \\
0 & 0 & \omega^2\end{array}\right)$.\\
\bottomrule
\end{tabular}
\caption{\label{tab:irrepTprime} The irreducible representation matrices for finite group $T'$.}
\end{table}

The nontrivial tensor product decomposition and Clebsch-Gordan (CG) coefficients of $T'$ (in the basis chosen in this work) are given by
\begin{align}
	\begin{pmatrix} x_{1}\\ x_{2} \end{pmatrix}_{\rep{2}} \otimes \begin{pmatrix} y_{1}\\ y_{2} \end{pmatrix}_{\rep{2} }&=\begin{pmatrix} x_{1}\\ x_{2} \end{pmatrix}_{\rep2'} \otimes \begin{pmatrix} y_{1}\\ y_{2} \end{pmatrix}_{\rep2''}=\left(x_{1}y_{2} - x_{2}y_{1}\right)_{\rep{1}} \oplus \begin{pmatrix} x_{1}y_{2} + x_{2}y_{1}\\ \sqrt{2}x_{2}y_{2}\\ -\sqrt{2}x_{1}y_{1} \end{pmatrix}_{\rep{3}}\;,\notag\\
	\begin{pmatrix} x_{1}\\ x_{2} \end{pmatrix}_{\rep2' } \otimes \begin{pmatrix} y_{1}\\ y_{2} \end{pmatrix}_{\rep2'}&=\begin{pmatrix} x_{1}\\ x_{2} \end{pmatrix}_{\rep{2}} \otimes \begin{pmatrix} y_{1}\\ y_{2} \end{pmatrix}_{\rep2''}=\left(x_{1}y_{2} - x_{2}y_{1}\right)_{\rep1''} \oplus \begin{pmatrix} \sqrt{2}x_{2}y_{2}\\ -\sqrt{2}x_{1}y_{1} \\ x_{1}y_{2} + x_{2}y_{1}  \end{pmatrix}_{\rep{3}}\;,\notag\\
	\begin{pmatrix} x_{1}\\ x_{2} \end{pmatrix}_{\rep2''} \otimes \begin{pmatrix} y_{1}\\ y_{2} \end{pmatrix}_{\rep2''}&=\begin{pmatrix} x_{1}\\ x_{2} \end{pmatrix}_{\rep{2}} \otimes \begin{pmatrix} y_{1}\\ y_{2} \end{pmatrix}_{\rep2'}=\left(x_{1}y_{2} - x_{2}y_{1}\right)_{\rep1'} \oplus \begin{pmatrix} -\sqrt{2}x_{1}y_{1} \\ x_{1}y_{2} + x_{2}y_{1}\\ \sqrt{2}x_{2}y_{2} \end{pmatrix}_{\rep{3}}\;,\notag\\
	\begin{pmatrix} x_{1}\\ x_{2} \end{pmatrix}_{\rep{2}} \otimes \begin{pmatrix} y_{1}\\ y_{2} \\ y_{3} \end{pmatrix}_{\rep{3}}&=\begin{pmatrix} x_{1} y_{1}+ \sqrt{2}x_{2} y_{3} \\ \sqrt{2} x_{1} y_{2} -x_{2} y_{1} \end{pmatrix}_{\rep{2}} \oplus \begin{pmatrix} x_{1} y_{2}+ \sqrt{2}x_{2} y_{1} \\ \sqrt{2}x_{1} y_{3} -x_{2} y_{2} \end{pmatrix}_{\rep2'} \oplus \begin{pmatrix} x_{1} y_{3}+ \sqrt{2}x_{2} y_{2} \\ \sqrt{2}x_{1} y_{1} -x_{2} y_{3} \end{pmatrix}_{\rep2''}\;,\notag\\
	\begin{pmatrix} x_{1}\\ x_{2} \end{pmatrix}_{\rep2'} \otimes \begin{pmatrix} y_{1}\\ y_{2} \\ y_{3} \end{pmatrix}_{\rep{3}}&=\begin{pmatrix} x_{1} y_{3}+ \sqrt{2}x_{2} y_{2} \\ \sqrt{2}x_{1} y_{1} -x_{2} y_{3} \end{pmatrix}_{\rep{2}} \oplus \begin{pmatrix} x_{1} y_{1}+ \sqrt{2}x_{2} y_{3} \\ \sqrt{2}x_{1} y_{2} -x_{2} y_{1} \end{pmatrix}_{\rep2'} \oplus \begin{pmatrix} x_{1} y_{2}+ \sqrt{2}x_{2} y_{1} \\ \sqrt{2}x_{1} y_{3} -x_{2} y_{2} \end{pmatrix}_{\rep2''}\;,\notag\\
	\begin{pmatrix} x_{1}\\ x_{2} \end{pmatrix}_{\rep2''} \otimes \begin{pmatrix} y_{1}\\ y_{2} \\ y_{3} \end{pmatrix}_{\rep{3}}&=\begin{pmatrix} x_{1} y_{2}+ \sqrt{2}x_{2} y_{1} \\ \sqrt{2}x_{1} y_{3} -x_{2} y_{2} \end{pmatrix}_{\rep{2}} \oplus \begin{pmatrix} x_{1} y_{3}+ \sqrt{2}x_{2} y_{2} \\ \sqrt{2}x_{1} y_{1} -x_{2} y_{3} \end{pmatrix}_{\rep2'} \oplus \begin{pmatrix} x_{1} y_{1}+ \sqrt{2}x_{2} y_{3} \\ \sqrt{2}x_{1} y_{2} -x_{2} y_{1} \end{pmatrix}_{\rep2''}\;,\notag\\
	\begin{pmatrix} x_{1}\\ x_{2} \\ x_{3} \end{pmatrix}_{\rep{3}} \otimes \begin{pmatrix} y_{1}\\ y_{2} \\ y_{3} \end{pmatrix}_{\rep{3}}&=\begin{pmatrix} x_{1} y_{1} + x_{2} y_{3} + x_{3} y_{2}  \end{pmatrix}_{\rep{1}} \oplus \begin{pmatrix} x_{1} y_{2} + x_{2} y_{1} + x_{3} y_{3}  \end{pmatrix}_{\rep1'} \oplus \begin{pmatrix} x_{1} y_{3} + x_{2} y_{2} + x_{3} y_{1}  \end{pmatrix}_{\rep1''}\notag\\
	&\qquad{}\oplus
	\begin{pmatrix} 2 x_{1} y_{1} - x_{2} y_{3} - x_{3} y_{2} \\ 2 x_{3} y_{3} - x_{1} y_{2} - x_{2} y_{1}  \\ 2 x_{2} y_{2}- x_{1} y_{3}  - x_{3} y_{1} \end{pmatrix}_{\rep{3}_\mathrm{S}} \oplus \begin{pmatrix} x_{3} y_{2}- x_{2} y_{3}  \\  x_{2} y_{1}- x_{1} y_{2}  \\ x_{1} y_{3} - x_{3} y_{1} \end{pmatrix}_{\rep{3}_\mathrm{A}}\;.\label{eq:Tprime_tensorProducts}
\end{align}
Here, the subscripts ``S'' and ``A'' denote symmetric and antisymmetric contractions, respectively.

%%%%%%%%%%%%%%%%%%%%%%%%%%%%%%%%%%%
A very common and useful quotient group for $T'$ is $A_4\cong\Z2^2\rtimes\Z3\cong T'/\Z2$ (GAP ID $[12,3]$).
$A_4$ has 12 elements, which can be generated by the generators $\mathrm S, \mathrm T$, satisfying the presentation
\begin{equation}
 A_4 ~=~ \left\langle \mathrm S, \mathrm T ~|~ \mathrm S^2 = (\mathrm S\mathrm T)^3 = \mathrm T^3 = \Id  \right\rangle\;.
\end{equation}
Its irreducible representations are a triplet $\rep3$ and three singlets $\rep1,\rep1',\rep1''$. Note that $T'$ 
is the double cover of $A_4$.

%%%%%%%%%%%%%%%%%%%%%%%%%%%%%%%%%%%%%%%%%%%%%%%%%%%%%
\subsection[Group Delta(54) = (54,8)]{\boldmath Group $\Delta(54)\cong [54,8]$ \unboldmath}
\label{app:Delta54}

The finite group $\Delta(54)\cong(\Z3 \times \Z3) \rtimes S_3 \cong[54,8]$ has 54 elements,
which can be generated by three generators $\mathrm A, \mathrm B$ and $\mathrm C$ satisfying the presentation
\begin{equation}
\label{eq:presenationDelta54}
\Delta(54)=\langle \mathrm A,\mathrm B,\mathrm C ~|~ \mathrm A^3=\mathrm B^3=\mathrm C^2=(\mathrm{AB})^3=(\mathrm{AB}^2)^3=(\mathrm{AC})^2=(\mathrm{BC})^2=\Id\rangle.
\end{equation}
Its irreducible representations are two singlets, four doublets and two triplets plus
their complex conjugates. The corresponding representation matrices are shown in \Cref{tab:irrepDelta54} below.
\begin{table}[th!]
\centering
\begin{tabular}{cccc}
\toprule
Irrep & $\rho_\mathbf{r}(\mathrm A)$ & $\rho_\mathbf{r}(\mathrm B)$  & $\rho_\mathbf{r}(\mathrm C)$  \\
\midrule
$\rep1$  & 1 & 1 & 1 \\
$\rep{1'}$  & 1 & 1 & -1 \\
\midrule 
$\rep2_1$ & $\left(\begin{array}{ll}1 & 0 \\ 0 & 1\end{array}\right)$ & $\left(\begin{array}{cc}\omega^2 & 0 \\ 0 & \omega\end{array}\right)$ & $\left(\begin{array}{ll}0 & 1 \\ 1 & 0\end{array}\right)$ \\
$\rep2_2$ & $\left(\begin{array}{cc}\omega^2 & 0 \\ 0 & \omega\end{array}\right)$ & $\left(\begin{array}{ll}1 & 0 \\ 0 & 1\end{array}\right)$ & $\left(\begin{array}{ll}0 & 1 \\ 1 & 0\end{array}\right)$ \\
$\rep2_3$ & $\left(\begin{array}{cc}\omega^2 & 0 \\ 0 & \omega\end{array}\right)$ & $\left(\begin{array}{cc}\omega^2 & 0 \\ 0 & \omega\end{array}\right)$ & $\left(\begin{array}{ll}0 & 1 \\ 1 & 0\end{array}\right)$ \\
$\rep2_4$  & $\left(\begin{array}{cc}\omega^2 & 0 \\ 0 & \omega\end{array}\right)$ & $\left(\begin{array}{cc}\omega & 0 \\ 0 & \omega^2\end{array}\right)$ & $\left(\begin{array}{ll}0 & 1 \\ 1 & 0\end{array}\right)$ \\ \midrule
$\rep3_1$ & $\left(\begin{array}{lll}0 & 1 & 0 \\ 0 & 0 & 1 \\ 1 & 0 & 0\end{array}\right)$ & $\left(\begin{array}{ccc}1 & 0 & 0 \\ 0 & \omega & 0 \\ 0 & 0 & \omega^2\end{array}\right)$ & $\left(\begin{array}{ccc}1 & 0 & 0 \\ 0 & 0 & 1 \\ 0 & 1 & 0\end{array}\right)$ \\
$\crep3_1$ & $\left(\begin{array}{lll}0 & 1 & 0 \\ 0 & 0 & 1 \\ 1 & 0 & 0\end{array}\right)$ & $\left(\begin{array}{ccc}1 & 0 & 0 \\ 0 & \omega^2 & 0 \\ 0 & 0 & \omega\end{array}\right)$ & $\left(\begin{array}{ccc}1 & 0 & 0 \\ 0 & 0 & 1 \\ 0 & 1 & 0\end{array}\right)$ \\
$\rep3_2$ & $\left(\begin{array}{lll}0 & 1 & 0 \\ 0 & 0 & 1 \\ 1 & 0 & 0\end{array}\right)$ & $\left(\begin{array}{ccc}1 & 0 & 0 \\ 0 & \omega & 0 \\ 0 & 0 & \omega^2\end{array}\right)$ & $\left(\begin{array}{ccc}-1 & 0 & 0 \\ 0 & 0 & -1 \\ 0 & -1 & 0\end{array}\right)$ \\
$\crep3_2$ & $\left(\begin{array}{lll}0 & 1 & 0 \\ 0 & 0 & 1 \\ 1 & 0 & 0\end{array}\right)$ & $\left(\begin{array}{ccc}1 & 0 & 0 \\ 0 & \omega^2 & 0 \\ 0 & 0 & \omega\end{array}\right)$ & $\left(\begin{array}{ccc}-1 & 0 & 0 \\ 0 & 0 & -1 \\ 0 & -1 & 0\end{array}\right)$ \\
\bottomrule
\end{tabular}
\caption{\label{tab:irrepDelta54} Irreducible representation matrices for the finite group $\Delta(54)$.}
\end{table}
It is useful to list the nontrivial tensor products of $\Delta(54)$ irreducible representations:
\begin{align}
\nonumber& \rep{1'} \otimes \rep{1'}=\rep{1},\quad
\rep{1'} \otimes \rep{2}_k=\rep{2}_k,\quad
\rep{1'} \otimes \rep{3}_1=\rep{3}_2,\quad
\rep{1'} \otimes \rep{3}_2=\rep{3}_1,\quad
\rep{1'} \otimes \crep{3}_1=\crep{3}_2,\quad
\rep{1'} \otimes \crep{3}_1=\crep{3}_2, \\
\nonumber& \rep2_k \otimes \rep2_k=\rep1 \oplus \rep{1'} \oplus \rep2_k,\quad
\rep2_k \otimes \rep2_{\ell}=\rep2_m \oplus \rep2_n \quad \text { with } k \neq \ell \neq m \neq n,\quad k, \ell, m, n=1, \ldots, 4, \\
\nonumber& \rep2_k \otimes \rep3_{\ell}=\rep3_1 \oplus \rep3_2, \quad
\rep2_k \otimes \crep3_{\ell}=\crep3_1 \oplus \crep3_2 \quad \text { for all } k=1, \ldots, 4,\quad \ell=1,2, \\
\nonumber& \rep3_{\ell} \otimes \rep3_{\ell}=\crep3_1 \oplus \crep3_1 \oplus \crep3_2,\quad
\rep3_1 \otimes \rep3_2=\crep3_2 \oplus \crep3_2 \oplus \crep3_1,\quad
\rep3_1 \otimes \crep3_1=\rep1 \oplus \rep2_1 \oplus \rep2_2 \oplus \rep2_3 \oplus \rep2_4, \\
\nonumber& \rep3_1 \otimes \crep3_2 = \rep{3}_2 \otimes \crep{3}_1 = \rep{1'} \oplus \rep2_1 \oplus \rep2_2 \oplus \rep2_3 \oplus \rep2_4,\\
& \rep3_2 \otimes \crep3_2=\rep1 \oplus \rep2_1 \oplus \rep2_2 \oplus \rep2_3 \oplus \rep2_4,\quad
\crep3_{\ell} \otimes \crep3_{\ell}=\rep3_1 \oplus \rep3_1 \oplus \rep3_2,\quad
\crep3_1 \otimes \crep3_2=\rep3_2 \oplus \rep3_2 \oplus \rep3_1 .
\end{align}
Some useful CG coefficients (in the basis chosen in this work) are as follows~\cite{Ishimori:2010au}
\begin{align}
	\begin{pmatrix} x_{1}\\ x_{2} \\ x_{3} \end{pmatrix}_{\rep{3}_\ell} \otimes \begin{pmatrix} y_{1}\\ y_{2} \\ y_{3} \end{pmatrix}_{\rep{3}_\ell}&=
	\begin{pmatrix} x_{1} y_{1} \\ x_{2} y_{2} \\ x_{3} y_{3}  \end{pmatrix}_{\crep{3}_1} \oplus \begin{pmatrix} x_{2} y_{3} + x_{3} y_{2} \\ x_{3} y_{1} + x_{1} y_{3} \\ x_{1} y_{2} + x_{2} y_{1} \end{pmatrix}_{\crep3_1} \oplus \begin{pmatrix} x_{2} y_{3} - x_{3} y_{2} \\ x_{3} y_{1} - x_{1} y_{3} \\ x_{1} y_{2} - x_{2} y_{1} \end{pmatrix}_{\crep3_2}\;,\quad \ell=1,2\notag\\
	\begin{pmatrix} x_{1}\\ x_{2} \\ x_{3} \end{pmatrix}_{\rep{3}_1} \otimes \begin{pmatrix} y_{1}\\ y_{2} \\ y_{3} \end{pmatrix}_{\crep{3}_2}&=
	(x_1 y_1 + x_2 y_2 + x_3 y_3)_{\rep{1'}} \oplus
	\begin{pmatrix} x_{1} y_{1} + \omega^2 x_2 y_2 + \omega x_3 y_3 \\ -\omega x_{1} y_{1} - \omega^2 x_2 y_2 - x_{3} y_{3}  \end{pmatrix}_{\rep{2}_1} \oplus
	\begin{pmatrix} x_{1} y_{2} + \omega^2 x_2 y_3 + \omega x_3 y_1 \\ -\omega x_{1} y_{3} - \omega^2 x_2 y_1 - x_{3} y_{2}  \end{pmatrix}_{\rep{2}_2} \notag\\
	&\oplus \begin{pmatrix} x_{1} y_{3} + \omega^2 x_2 y_1 + \omega x_3 y_2 \\ -\omega x_{1} y_{2} - \omega^2 x_2 y_3 - x_{3} y_{1}  \end{pmatrix}_{\rep{2}_3} \oplus
	\begin{pmatrix} x_{1} y_{3} + x_2 y_1 + x_3 y_2 \\ - x_{1} y_{2} - x_2 y_3 - x_{3} y_{1}  \end{pmatrix}_{\rep{2}_4}\;,\notag\\
	\begin{pmatrix} x_{1}\\ x_{2} \\ x_{3} \end{pmatrix}_{\crep{3}_1} \otimes \begin{pmatrix} y_{1}\\ y_{2} \\ y_{3} \end{pmatrix}_{\rep{3}_2}&=
	(x_1 y_1 + x_2 y_2 + x_3 y_3)_{\rep{1'}} \oplus
	\begin{pmatrix} x_{1} y_{1} + \omega^2 x_2 y_2 + \omega x_3 y_3 \\ -\omega x_{1} y_{1} - \omega^2 x_2 y_2 - x_{3} y_{3}  \end{pmatrix}_{\rep{2}_1} \oplus
	\begin{pmatrix} x_{1} y_{3} + \omega^2 x_2 y_1 + \omega x_3 y_2 \\ -\omega x_{1} y_{2} - \omega^2 x_2 y_3 - x_{3} y_{1}  \end{pmatrix}_{\rep{2}_2} \notag\\
	&\oplus \begin{pmatrix} x_{1} y_{2} + \omega^2 x_2 y_3 + \omega x_3 y_1 \\ -\omega x_{1} y_{3} - \omega^2 x_2 y_1 - x_{3} y_{2}  \end{pmatrix}_{\rep{2}_3} \oplus
	\begin{pmatrix} x_{1} y_{2} + x_2 y_3 + x_3 y_1 \\ - x_{1} y_{3} - x_2 y_1 - x_{3} y_{2}  \end{pmatrix}_{\rep{2}_4}
	\;.
	\label{eq:Delta54_tensorProducts}
\end{align}

%%%%%%%%%%%%%%%%%%%%%%%%%%%%%%%%%%%%%%%%%%%%%%%%%%%%%%%%%%%%%%%%%%%%%%%%%%
%  Bibliography
%%%%%%%%%%%%%%%%%%%%%%%%%%%%%%%%%%%%%%%%%%%%%%%%%%%%%%%%%%%%%%%%%%%%%%%%%%
%\bibliography{Orbifold}
%\bibliographystyle{NewArXiv}
\providecommand{\bysame}{\leavevmode\hbox to3em{\hrulefill}\thinspace}
\frenchspacing
\newcommand{\origttfamily}{}
\let\origttfamily=\ttfamily
\renewcommand{\ttfamily}{\origttfamily \hyphenchar\font=`\-}

%%%%%%%%%%%%%%%%%%%%%%%%%%%%%%%%%%%%%%%%%%%%%%%%%%%%%%%%%%%%%%%%%%%%%%%%%%
%  Acronyms
%%%%%%%%%%%%%%%%%%%%%%%%%%%%%%%%%%%%%%%%%%%%%%%%%%%%%%%%%%%%%%%%%%%%%%%%%%

\begin{acronym}
  \acro{2HDM}{two-Higgs doublet model}  
  \acro{BSM}{beyond the standard model}
  \acro{BU}{bottom-up}
  \acro{EFT}{effective field theory}
  \acro{FCNC}{flavor changing neutral current}
  \acro{FI}{Fayet--Iliopoulos \cite{Fayet:1974jb}} 
  \acro{GS}{Green--Schwarz \cite{Green:1984sg}}
  \acro{GUT}{Grand Unified Theory}
  \acro{IO}{inverted ordering}
  \acro{IR}{infrared}
  \acro{LEET}{low-energy effective theory}
  \acro{LHC}{Large Hadron Collider}
  \acro{MIHO}{modular invariant holomorphic observables}
  \acro{MSSM}{minimal supersymmetric standard model}
  \acro{NO}{normal ordering}
  \acro{OPE}{operator product expansion}
  \acro{QFT}{quantum field theory}
  \acro{RG}{renormalization group}
  \acro{RGE}{renormalization group equation}
  \acro{SM}{standard model}
  \acro{SUSY}{supersymmetry}
  \acro{TD}{top-down}
  \acro{UV}{ultraviolet}
  \acro{VEV}{vacuum expectation value}
  \acro{VVMF}{vector-valued modular form}
  \acro{CFT}{conformal field theory}
\end{acronym}

\end{document}